\documentstyle[amssymb,aps,epsfig,multicol]{revtex}

\begin{document}

\title{Spatiotemporal Communication with Synchronized Optical Chaos}

\author{J. Garc\'{\i}a-Ojalvo}
\address{Departament de F\'{\i}sica i Enginyeria Nuclear,
Universitat Polit\`{e}cnica de Catalunya, Colom 11,
E-08222 Terrassa, Spain}
\author{R. Roy}
\address{Department of Physics and Institute for Physical Science and
Technology\\
University of Maryland, College Park, Maryland 20742}

\maketitle

\begin{abstract}
We propose a model system that allows communication of spatiotemporal
information using an optical chaotic carrier waveform. The system is based on 
broad-area nonlinear optical ring cavities, which exhibit spatiotemporal chaos
in a wide parameter range. Message recovery is possible through chaotic
synchronization between transmitter and receiver. Numerical simulations
demonstrate the feasibility of the proposed scheme, and the benefit of the
parallelism of information transfer with optical wavefronts.

\end{abstract}


\begin{multicols}{2}
{
One of the most appealing applications of chaos is the possibility of
using chaotic signals as broad-band carriers of information, which could
lead to a simple implementation of spread-spectrum communication systems.
Many of the schemes devised so far to that end are based on the occurrence 
of synchronization between two chaotic systems \cite{pecora90}.
Following the original implementation of this approach in electronic circuits
\cite{kocarev92,cuomo93}, special attention has been paid to using optical
systems \cite{pere}, which offer the possibility of high-speed data transfer
in all-optical communication systems \cite{claudio}.
Optical chaotic communication was
recently demonstrated in fiber lasers \cite{greg}, diode lasers
with external nonlinearity \cite{goedgebuer}, and diode lasers with
optical feedback \cite{shore}.

In most optical realizations of communications with chaos, the message
to be encoded drives the nonlinear transmitter, so that the message
and carrier become mixed in a nontrivial way. The resulting output is
injected into a receiver, which, upon synchronization to the transmitter,
allows for recovery of the original signal. The optical schemes developed
so far have used
purely temporal chaotic signals as information carriers. The present
Letter proposes a nonlinear optical device exhibiting spatiotemporal
chaos as the basis of a communication system capable of transmitting
information {\em in space and time}.
Chaotic behavior in spatial degrees of freedom has recently been used
for multichannel communication with multimode semiconductor lasers
\cite{white99}. In that case, however, only variations of the electric
field along its propagation direction were considered.
Information was encoded in the different longitudinal cavity modes,
demonstrating a technique for multiplexing.
Spatiotemporal communication, on the other hand, utilizes
the inherent large scale parallelism of information transfer that is
possible with {\em broad-area} optical wavefronts. As in the previous
cases \cite{henry},
our scheme requires the existence of synchronization between transmitter
and receiver. Synchronization of spatiotemporal chaos has been investigated
extensively in previous years, but most studies have been restricted
to nonlinear oscillator arrays \cite{kocarev96}, coupled map lattices
\cite{xiao96} and model partial differential equations
\cite{amengual97,kocarev97,boccaletti99}. Our system, on the other hand,
is represented by an infinite-dimensional map spatially coupled in a continuous
way by light diffraction.
\begin{figure}
\centerline{
\epsfig{file=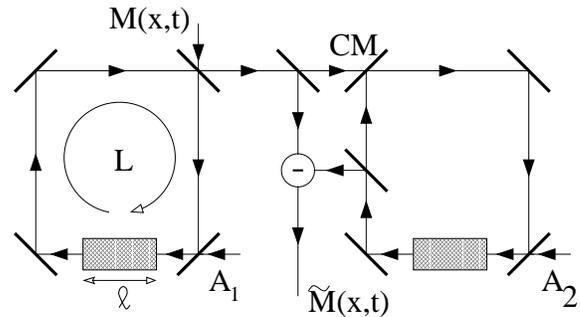,width=75mm}
}
\noindent
\begin{minipage}{0.48\textwidth}
\caption{Scheme for communicating spatiotemporal information using
optical chaos. CM is a coupling mirror.}
\label{fig:setup}
\end{minipage}
\end{figure}
The experimental setup is shown schematically in Fig.~\ref{fig:setup}.
Two optical ring cavities are unidirectionally coupled by a light beam
extracted from the left ring (the transmitter) and partially injected
into the right one (the receiver). Each
cavity contains a broad-area nonlinear absorbing medium, and is subject
to a continuously injected plane wave $A_i$. Light diffraction will be
taken into account during propagation through the medium, in such a way
that a nonuniform distribution of light in the plane transverse to the
propagation direction may appear. In fact, an infinite number of transverse
modes will in principle be allowed to oscillate within the cavity.
A spatiotemporal message $M$ can be introduced into the transmitter's
cavity, and recovered as $\widetilde{M}$ in the receiver as explained below.

When no message is introduced and the receiver is absent,
the transmitter is a standard nonlinear ring cavity,
well known to exhibit
temporal optical chaos \cite{ikeda79}. When
transverse effects due to light diffraction are taken into account,
a rich variety of spatiotemporal instabilities appear \cite{mclaughlin85},
giving rise
to solitary waves \cite{mclaughlin83}, period-doubling bifurcations
\cite{haeltermann93}, spatial patterns \cite{leberre96}, and
spatiotemporal chaos \cite{sauer96,leberre97}. This latter behavior is the
one in which we are interested, since such chaotic waveforms will be used
as information carriers in our setup.

The propagation of light through the nonlinear medium can
be described by the following equation for the slowly-varying
complex envelope $E_n(\vec x,z)$ of the electric field (assumed to be
linearly polarized) in the n-th passage through the resonator \cite{sauer96}:
\begin{equation}
\frac{\partial E_n(\vec x,z)}{\partial t}=
\frac{i}{2k}\nabla^2E_n(\vec x,z)-\frac{\alpha(1+i\Delta)}
{1+4|E_n|^2}\;E_n(\vec x,z)\,.
\label{eq:prop}
\end{equation}
The first term on the right-hand side of (\ref{eq:prop}) describes
diffraction, and the second saturable absorption. The propagation direction
is denoted by $z$, whereas $\vec x$ is a vector in the plane orthogonal
to the propagation direction. Equation (\ref{eq:prop}) obeys the 
boundary condition
\begin{equation}
E_n(\vec x,0)=\sqrt{T}A+R\,\exp(ikL)E_{n-1}(\vec x,\ell)\,,
\label{eq:map}
\end{equation}
which corresponds to the infinite-dimensional map that is the object of
our analysis. $z=0$ in (\ref{eq:map}) denotes the input of the nonlinear
medium, which has length $\ell$. The total length of the cavity is $L$.
Other parameters of the model are the absorption coefficient $\alpha$
of the medium, the detuning $\Delta$ between the atomic transition and
cavity resonance frequencies, the transmittivity $T$ of the input mirror,
and the total return coefficient $R$ of the cavity (fraction of light
intensity remaining in the cavity after one round trip). The injected
signal, with amplitude $A$ and wavenumber $k$, is taken to
be in resonance with a longitudinal cavity mode.

Previous studies have shown that for $\Delta<0$, model
(\ref{eq:prop})-(\ref{eq:map}) exhibits irregular dynamics
in both space and time for $A$ large enough \cite{sauer96}. This
spatiotemporally chaotic behavior can become synchronized to that of
a second cavity, also operating in a chaotic regime, coupled to the
first one as shown in Fig. \ref{fig:setup}. The coupling mechanism
can be modeled in the following form \cite{henry}:
\begin{eqnarray}
& &E_n^{(1)}(\vec x,0)={\cal F}^{(1)}\left[E_{n-1}^{(1)}(\vec x,\ell)\right] 
\nonumber
\\
& &E_n^{(2)}(\vec x,0)={\cal F}^{(2)}\left[(1-c)E_{n-1}^{(2)}(\vec x,\ell)
+cE_{n-1}^{(1)}(\vec x,\ell)\right]\,, 
\label{eq:sync}
\end{eqnarray}
where the application ${\cal F}^{(i)}$ represents the action of the 
map (\ref{eq:map}) in every round trip. The coupling coefficient $c$ is
given by the transmittivity of the coupling mirror CM (Fig. \ref{fig:setup}).
The superindices 1 and 2 represent the transmitter and receiver, respectively.
Earlier studies have shown that local sensor coupling is enough to achieve
synchronization of spatiotemporal chaos in model continuous
equations \cite{junge00}.
In our optical model, however, the whole spatial domain can be coupled to
the receiver in a natural way.

To estimate the synchronization efficiency of scheme (\ref{eq:sync}),
we have evaluated the synchronization error \cite{kocarev97}
${\displaystyle
e_n=\sqrt{\frac{1}{S}\int_S \left|E_n^{(1)}(\vec x,\ell)-
E_n^{(2)}(\vec x,\ell)\right|^2\,d\vec x}}$,
where $S$ is the size of the system. This quantity has been computed
for increasing values of the coupling coefficient $c$, by numerically
integrating model (\ref{eq:prop})-(\ref{eq:map}) for both transmitter
and receiver operating in a regime of spatiotemporal chaos, and using
the coupling scheme (\ref{eq:sync}). Simulations have been performed
in a 1-d lattice of 1000 cells of size $dx=0.1$ spatial units, using
a pseudospectral code for the propagation equation (\ref{eq:prop}).
Similar parameter values to those of Ref. \cite{sauer96} are used here.
The initially uncoupled systems evolve in time starting from
arbitrary initial conditions, and after 100 round trips, when their
unsynchronized chaotic dynamics is fully developed, coupling is
switched on. The synchronization error $e_n$ is measured 100 round trips
later.  The results are shown in Fig. \ref{fig:sync}(a), which plots the
value of $e_n$ for increasing coupling strengths. According to these
results, a high degree of synchronization can be obtained for couplings
as low as 40\%.

\begin{figure}[htb]
\centerline{
\epsfig{file=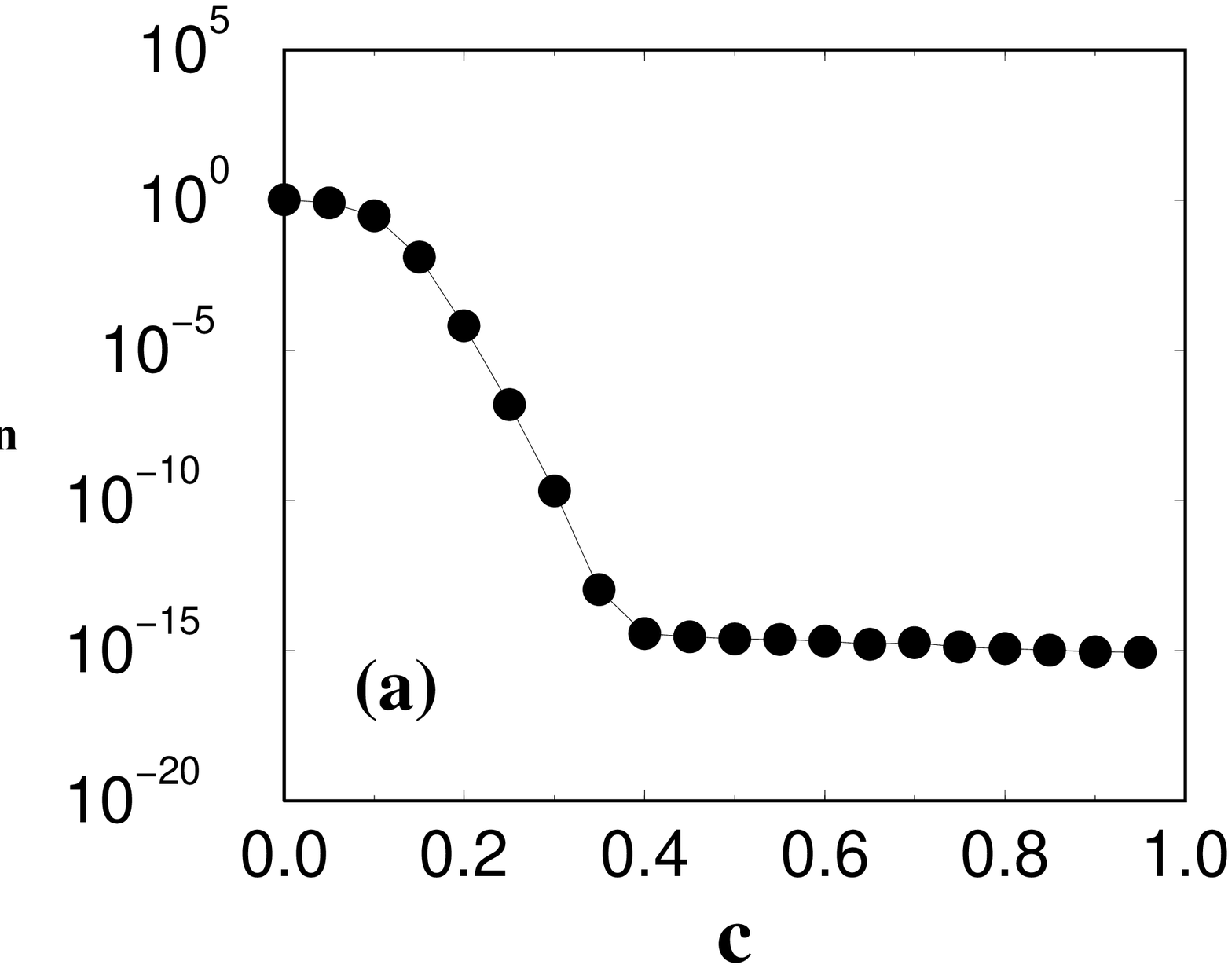,width=41mm}
\epsfig{file=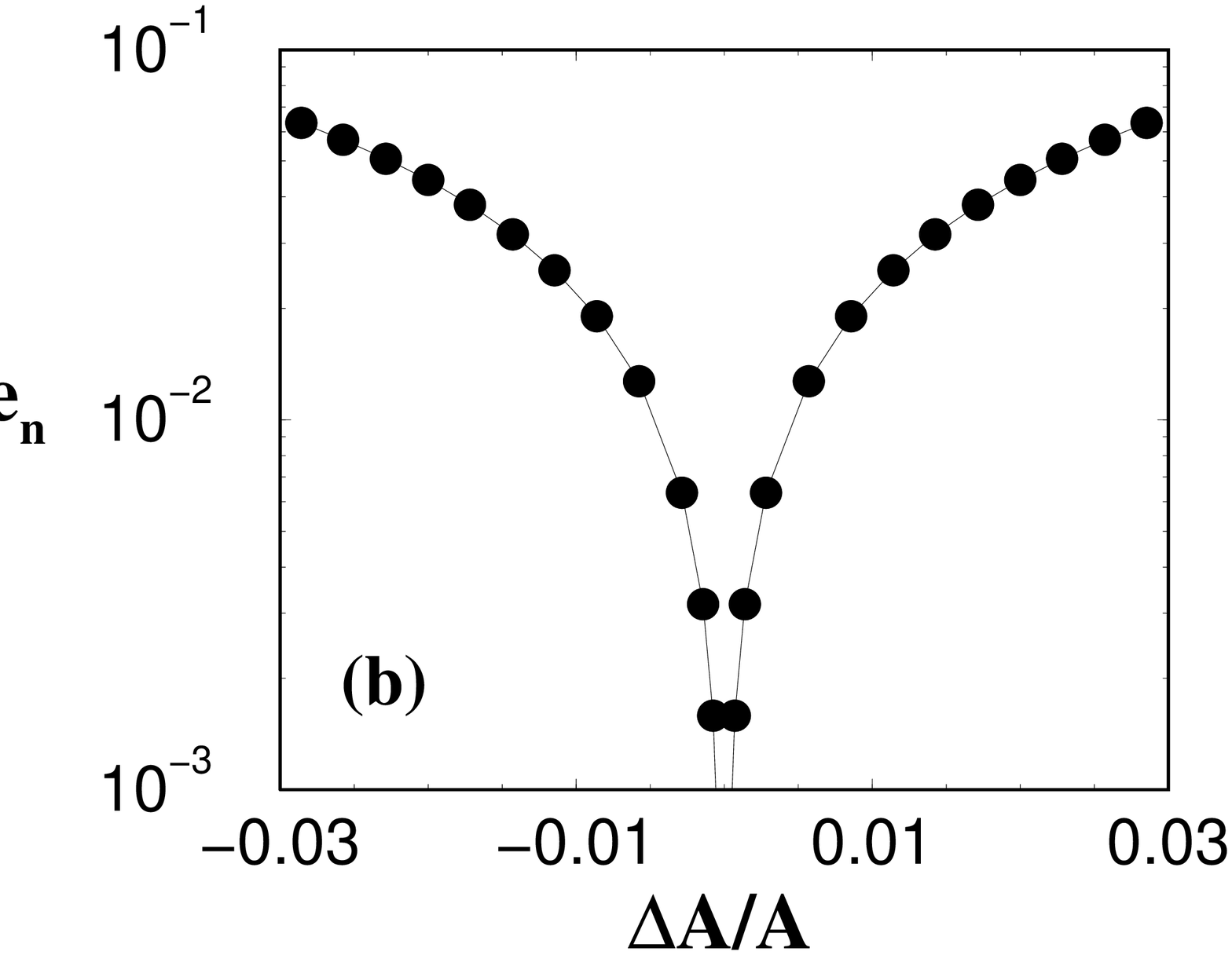,width=41mm}
}
\noindent
\begin{minipage}{0.48\textwidth}
\caption{Synchronization error $e_n$ versus (a) coupling coefficient
and (b) injected-amplitude mismatch. Parameters common to the
two cavities are $\alpha=100.0$, $\Delta=-10.0$, $R=0.9$, $T=0.1$,
$k=100.0$, $\ell=0.01$, and $L=0.015$. In (a), the common value of
$A$ is $7.0$, which is also the value used for the transmitter in (b).
}
\label{fig:sync}
\end{minipage}
\end{figure}

Another important issue to address at this point is how sensitive
synchronization is to differences between the two cavities. We have
extensively analyzed the effect of different parameter mismatches
on the synchronization error $e_n$. Our
results indicate that parameters such as the absorption coefficient
$\alpha$, the detuning $\Delta$ and the nonlinear medium length $\ell$
can be varied as much as 50\% and still keep $e_n$ below $10^{-2}$.
More sensitive parameters are the total length $L$ of the cavity [due
to its appearance in the phase-change term of (\ref{eq:map})] and the
amplitude $A$ of the injected signal. Since the two cavity lengths can
be matched experimentally, we now examine in detail the effect of
a mismatch in $A$. This parameter could be controlled in real time
if necessary, and hence act as a control parameter for synchronization.
The variation of $e_n$ versus relative mismatch of $A$ is shown in
Fig. \ref{fig:sync}(b). It can be seen that synchronization is quickly
degraded as the two injected amplitudes differ, with $e_n$ increasing
well above $10^{-2}$ for mismatches of the order of 1\%. Therefore, the
value of $A$ is critical for obtaining synchronization in the system.

We now use the synchronizing system described above to encode and decode
information in space and time using the spatiotemporal chaotic carrier.
We modify the scheme of Eqs. (\ref{eq:sync}) according to Fig.
\ref{fig:setup}, which leads to:
\begin{eqnarray}
& &E_n^{(1)}(\vec x,0)={\cal F}^{(1)}\left[E_{n-1}^{(1)}(\vec x,\ell)
+M_{n-1}(\vec x)\right] 
\nonumber
\\
& &E_n^{(2)}(\vec x,0)={\cal F}^{(2)}\left[(1-c)E_{n-1}^{(2)}(\vec x,\ell)
\right.
\nonumber
\\
& &
\qquad\qquad\qquad\qquad
\left.
+c\left(E_{n-1}^{(1)}(\vec x,\ell)+M_{n-1}(\vec x)\right)\right]\,, 
\label{eq:encod}
\end{eqnarray}
If synchronization between transmitter and receiver is achieved,
it will be possible to decode the signal by simply subtracting the
transmitted signal and the one in the receiver:
$\widetilde{M}_n(\vec x)=E_n^{(1)}(\vec x,\ell)+M_n(\vec x)-
E_n^{(2)}(\vec x,\ell)$.
In the case of no mismatch, it can be seen analytically in a
straightforward way that, as the coupling coefficient $c$ tends to 1,
the difference $|E_n^{(1)}-E_n^{(2)}|\to 0\quad \forall \vec x$,
which corresponds to perfect synchronization, and hence to perfect
message recovery.
It should be noted that the message is not merely added to the
chaotic carrier, but rather the former is {\em driving} the nonlinear
transmitter itself. Therefore, as we will see in what follows, the
amplitude of the message need not be much smaller than that of the
chaotic signal to provide good masking of the information.

The scheme described above has been tested by encoding an analog 1-d
signal with complex evolution in space and time. The sample signal
chosen is the spectrogram of a sample of speech. Chaotic systems have
been used in the past to encode speech waveforms \cite{cuomo93,henry},
but the information that such signals provide is insufficient for
voice-recognition
purposes. Spectrograms, on the other hand, contain information on a
broad range of frequencies as time evolves. Figure
\ref{fig:spect}(a) shows a grayscale spectrogram of the word ``compute'',
with frequency components in the horizontal axis and time evolving from
bottom to top. We will encode the frequency information in the 1-d
transverse direction of our setup. The real part of the transmitted signal
is shown in Fig. \ref{fig:spect}(b) for a message amplitude maximum of
0.5. This value should be compared to the maximum intensity of the chaotic
carrier, which oscillates between 1 and 10, approximately, for the
parameters chosen.
The spatiotemporal chaotic state of the signal can be clearly observed.
Finally, Fig. \ref{fig:spect}(c) shows the detected message,
for a 90\% coupling between transmitter and receiver.

Figure \ref{fig:spect} qualitatively shows that, even though coupling between
transmitter
and receiver is not complete, information varying in time {\em and} space
can be successfully transmitted and recovered with the setup described in
Fig. \ref{fig:setup}. In order to have a quantitative measure of this
effect, we have estimated the mutual information between input and
output message signals, and its dependence on several system parameters.
To that end, we discretize the values of $M$ and $\widetilde{M}$ in space-time
points, and compute the probability distributions $p(x)$, $p(y)$,
and the joint probability $p(x,y)$, where $x$ and $y$ are the different
values that $M$ and $\widetilde{M}$ may take, respectively. A measure of
the mutual information between the two sets of data is given by
${\displaystyle I=-\sum_{x,y}p(x,y)\ln [p(x)p(y)/p(x,y)]}$, where
the sums run over all
possible values of $M$ and $\widetilde{M}$. This mutual information
function is 0 for completely independent data sets, and equal to
the entropy of the common signal, $H=-\sum_xp(x)\ln p(x)$,
when the two messages are identical.
Figure \ref{fig:ixy}(a) shows the value of
the mutual information $I$, for the message encoding proposed in
Fig. \ref{fig:spect}, versus the coupling coefficient $c$. It can be seen
that, as $c$ increases, $I$ grows from 0 to perfect recovery, corresponding
to the entropy of the input image, given by the horizontal dashed line
in the figure. This result shows that, even though good synchronization
appears for $c\gtrsim 0.4$, satisfactory message recovery requires
coupling coefficients closer to unity. This can also be seen by examining
the behavior of the entropy $H$ of the recovered image, plotted as empty
squares in Fig. \ref{fig:ixy}(a): for values of $c$ substantially smaller
than 1, the entropy of the recovered data is appreciably larger than that
of the input message, indicating a higher degree of randomness in the
former. Finally, the behavior of the mutual information in the presence
of noise is shown as empty diamonds in Fig. \ref{fig:ixy}(a).
Uncorrelated, uniformly distributed noise is added continuously to the
communication channel, with
an amplitude 1\% that of the message. The results show
that the scheme is reasonably robust, in agreement with previous studies
\cite{white99}. A more systematic analysis of this issue is in progress.

\begin{figure}[htb]
\centerline{
\epsfig{file=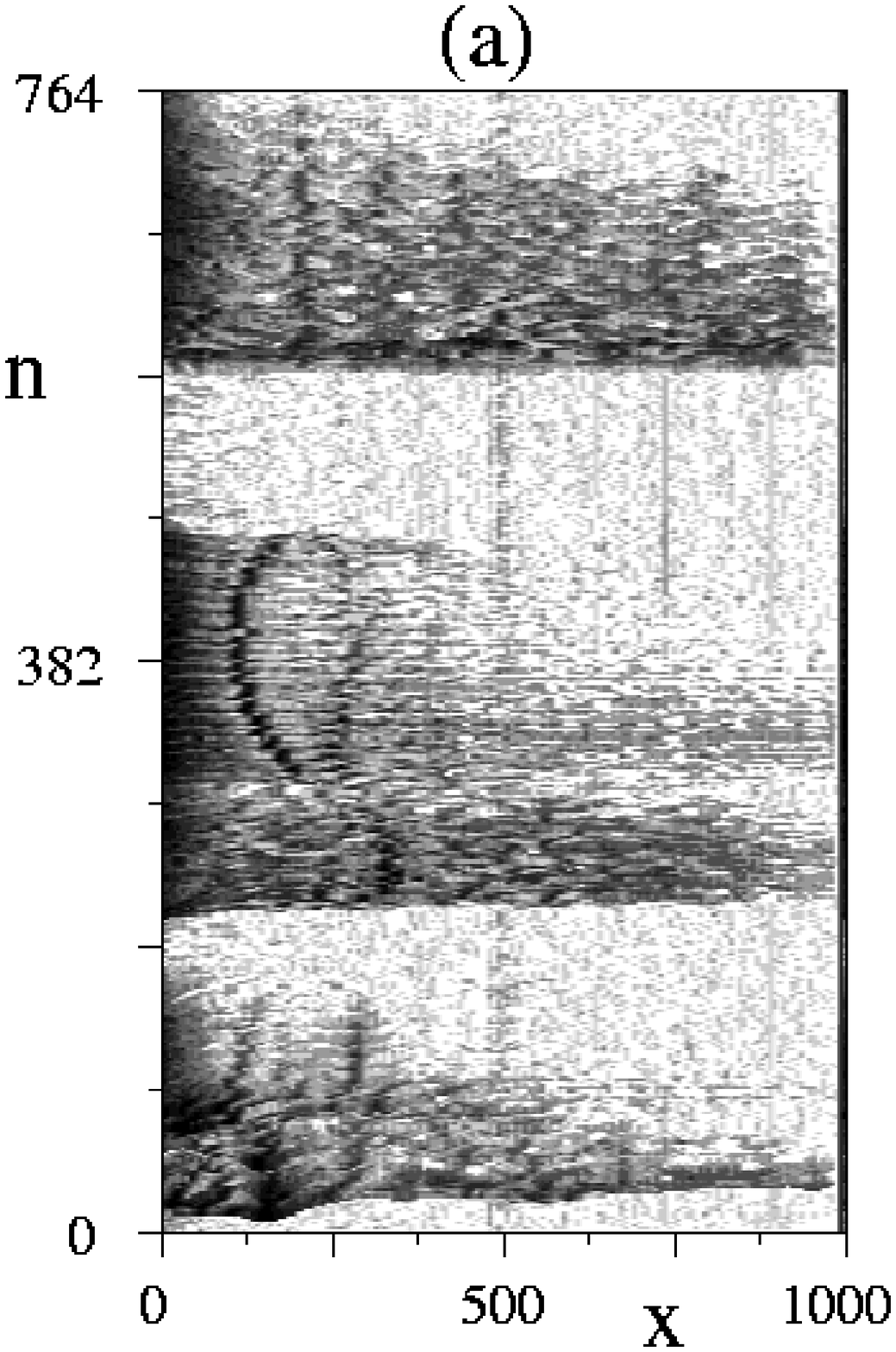,width=28mm}
\epsfig{file=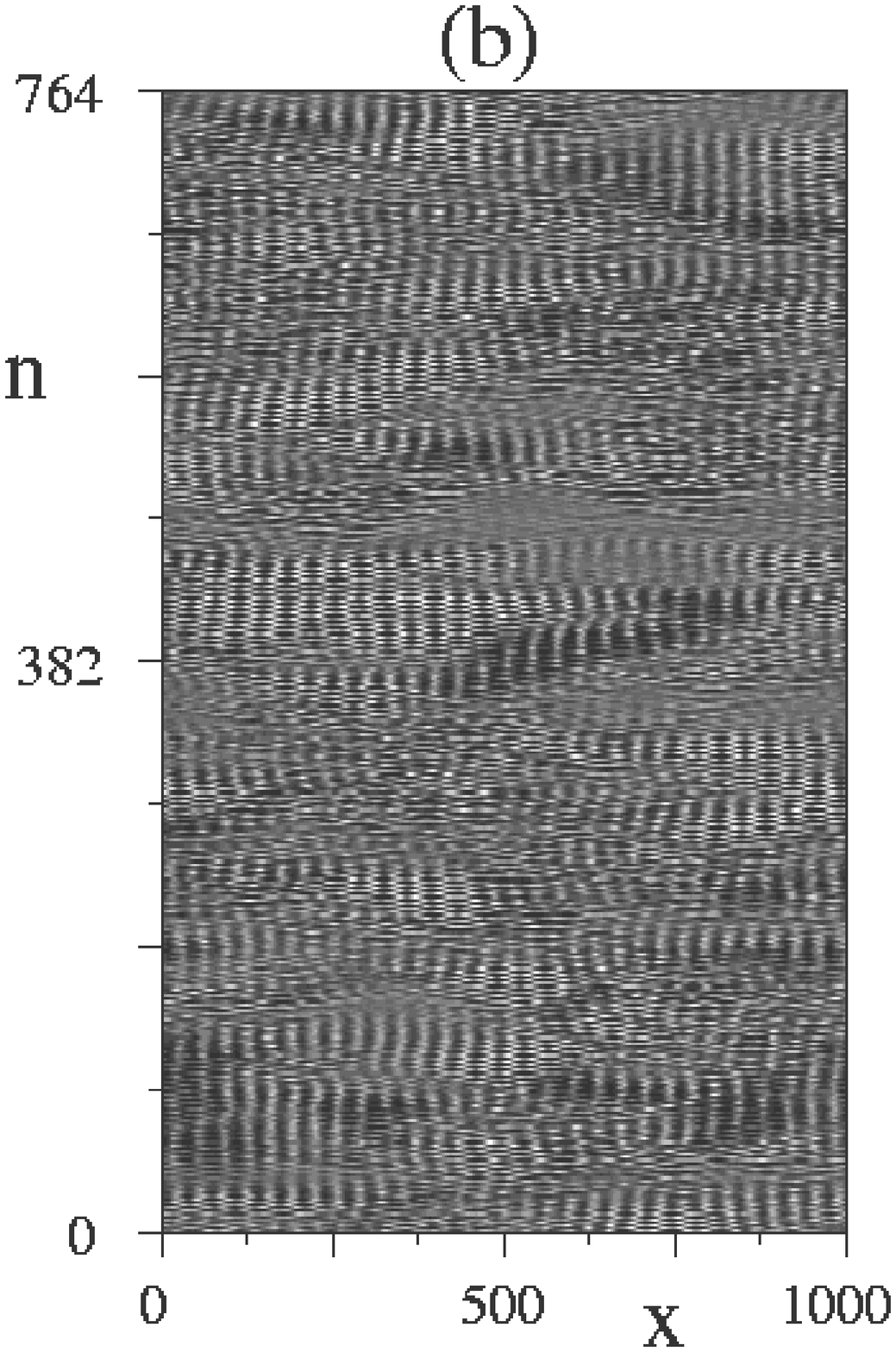,width=28mm}
\epsfig{file=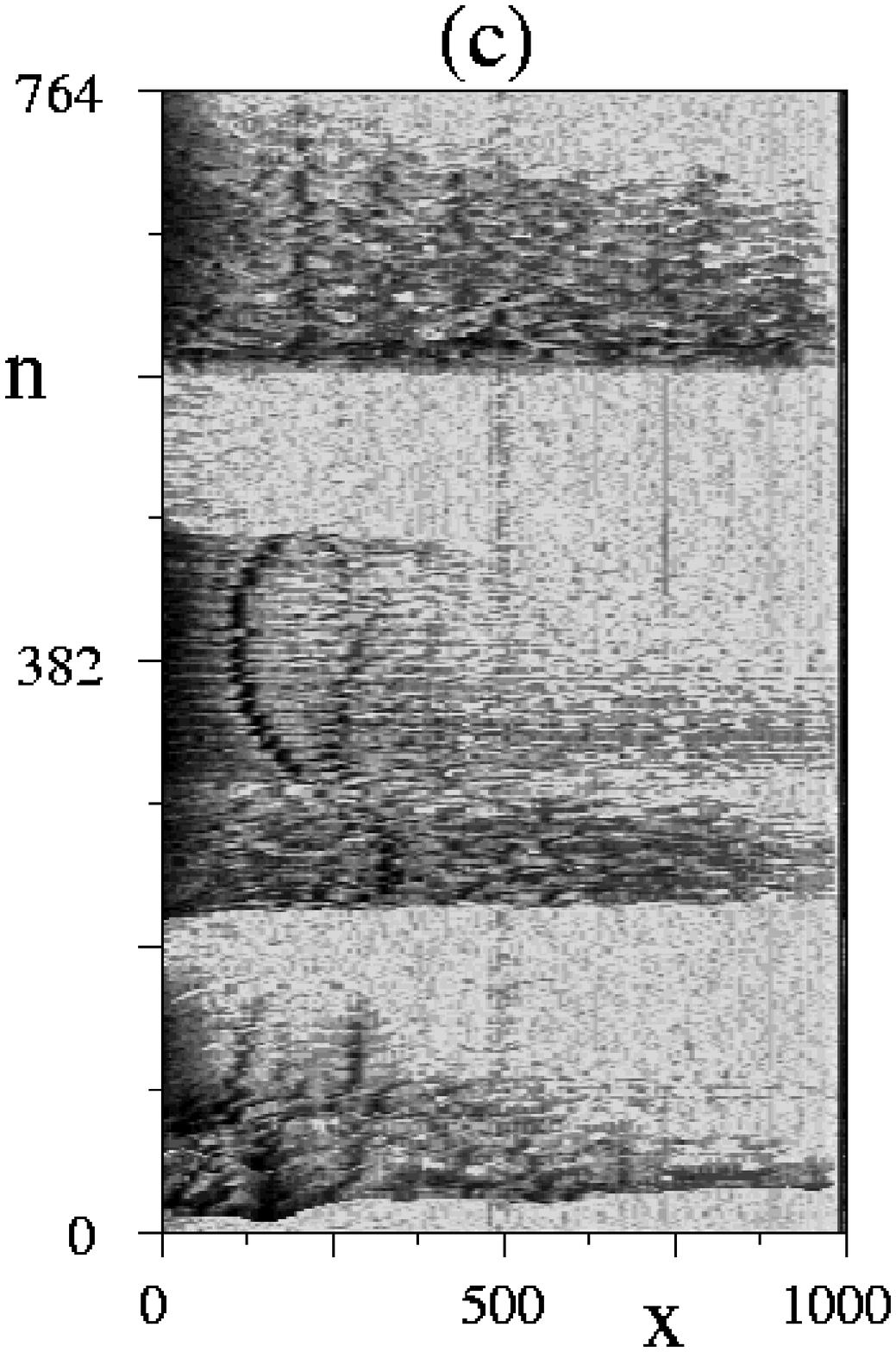,width=28mm}
}
\noindent
\begin{minipage}{0.48\textwidth}
\caption{Transmission of 1-d spatiotemporal data. (a) Input spectrogram;
(b) real part of transmitted signal; (c) recovered data.
Parameters are those Fig. \protect\ref{fig:sync}(a), plus $c=0.9$.
}
\label{fig:spect}
\end{minipage}
\end{figure}

We have also examined the effect of parameter mismatch on the efficiency
of message recovery. As in the synchronization characterization,
we have concentrated on the influence of the most sensitive parameter,
namely the amplitude $A$ of the injected signal. The data plotted in
Fig. \ref{fig:ixy}(b) show that a slight mismatch in the value of $A$
will degrade
recovery, by leading to values of $I$ much smaller that the entropy
of the input message, and to a recovered message with substantially
larger entropy than the original.

\begin{figure}[htb]
\centerline{
\epsfig{file=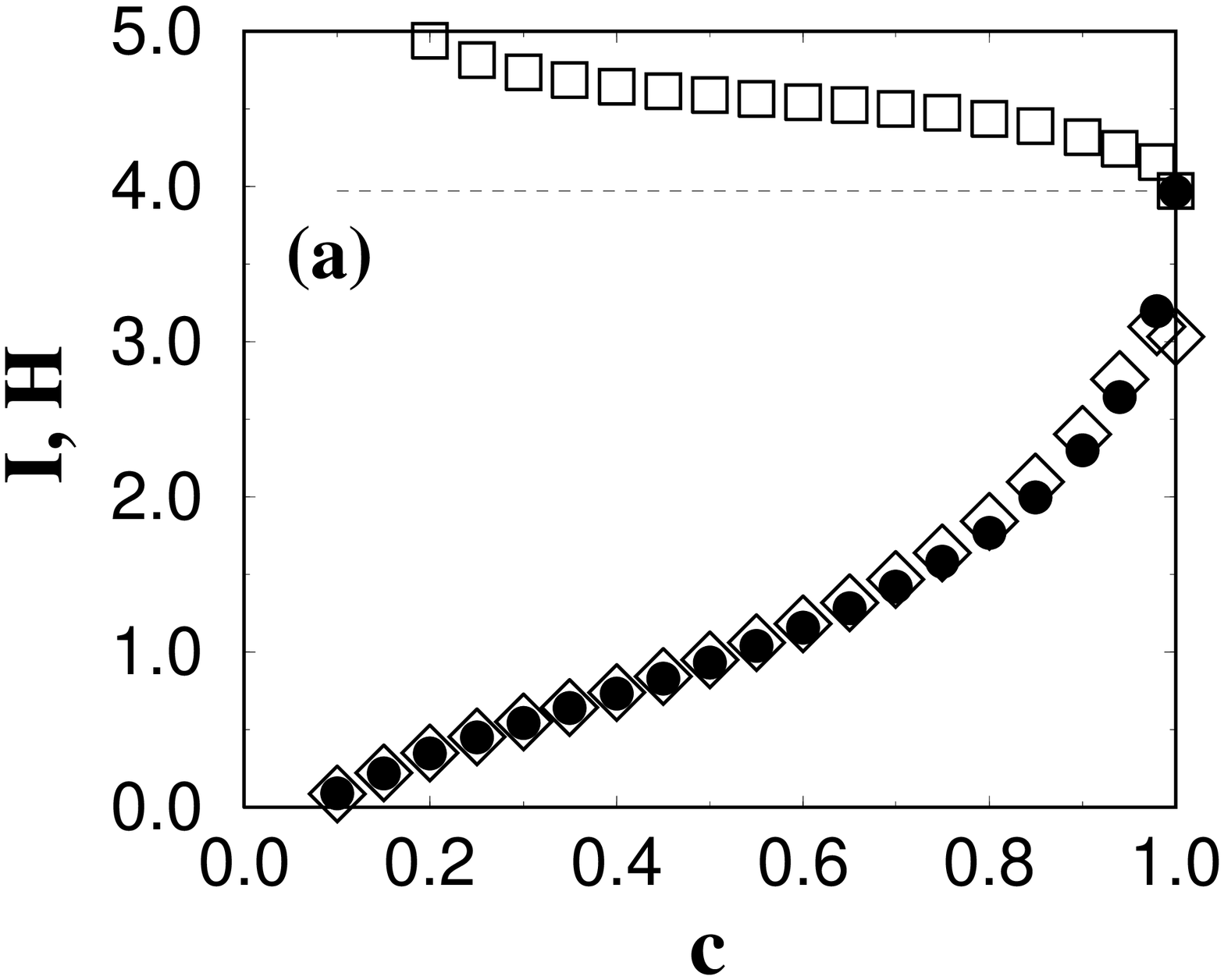,width=41mm}
\epsfig{file=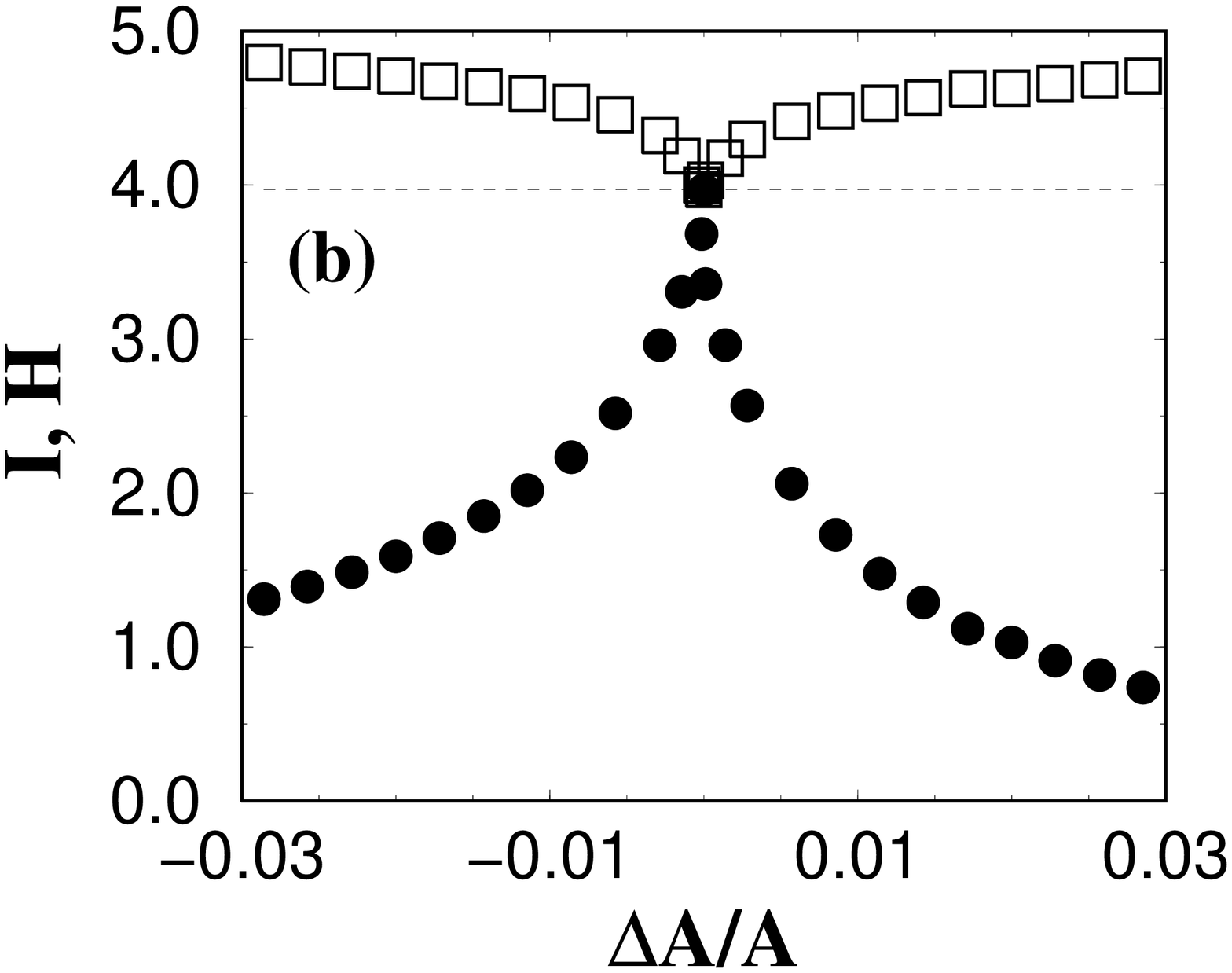,width=41mm}
}
\noindent
\begin{minipage}{0.48\textwidth}
\caption{Information measures corresponding to the message encoding
of Fig. \protect\ref{fig:spect}. Full circles: mutual information $I$;
empty squares: entropy $H$ of the recovered data; horizontal dashed
line: entropy of the original image. Empty diamonds are the values of
$I$ in the presence of noise (see text).
Parameters are those of Fig. \protect\ref{fig:sync}.
}
\label{fig:ixy}
\end{minipage}
\end{figure}

Finally, we should note that our setup is also suitable for the
transmission of two-dimensional information. To illustrate this,
we have chosen to encode a static 2-d image with the same mechanism
discussed above. Figure \ref{fig:bird} shows the results obtained
in this case. As in Fig. \ref{fig:spect}, the left plot depicts the
input message, the middle plot the real part of the transmitted signal
(a snapshot of it, in this case), and the right plot the recovered
data. The message amplitude maximum is now 0.01.
Simulations are now performed on a square array with 256$\times$256
pixels of width $dx=1.0$. The image is clearly recognizable
even though the coupling coefficient is now as low as 0.7.

\begin{figure}[htb]
\centerline{
\epsfig{file=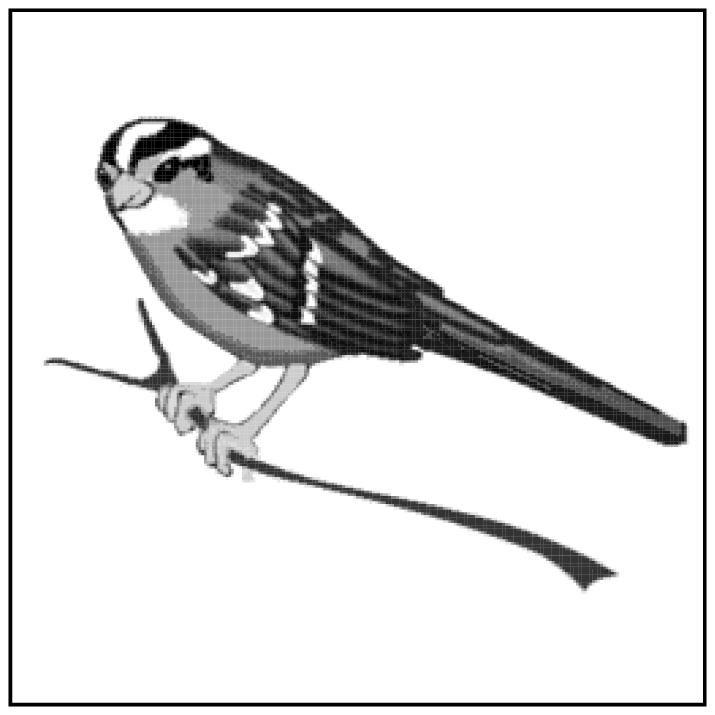,width=27mm}
\epsfig{file=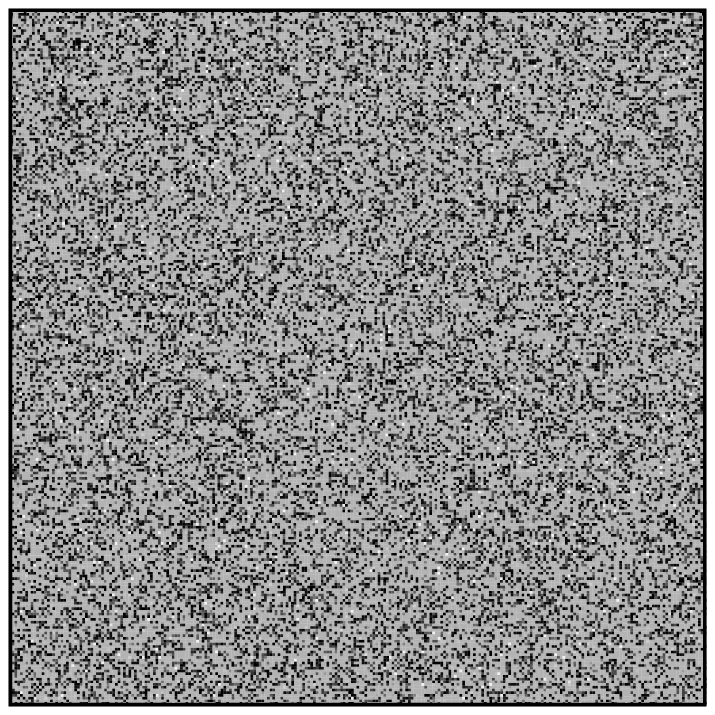,width=27mm}
\epsfig{file=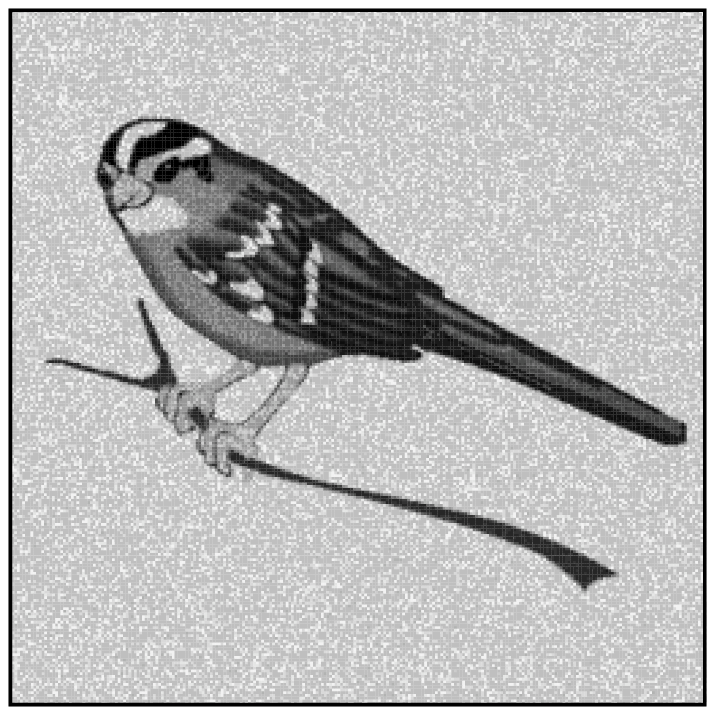,width=27mm}
}
\noindent
\begin{minipage}{0.48\textwidth}
\caption{Transmission of a 2-d spatiotemporal static image.
(a) Input image; (b) real part of the transmitted
signal at a certain time; (c) recovered data.
Parameters are those Fig. \protect\ref{fig:sync}(a), plus $c=0.7$.
}
\label{fig:bird}
\end{minipage}
\end{figure}

In conclusion, we have proposed a nonlinear optical model system that
allows encoding and decoding information in space and time by means of
spatiotemporal chaos synchronization. Synchronization occurs for
a wide range of coupling values and system parameters. Spatiotemporal
information can be successfully recovered for large enough coupling
between transmitter and receiver, and for small enough parameter mismatches.
The proposed
setup could be experimentally implemented upon identification of a
suitable broad-area nonlinear medium.

Financial support from NATO (project CRG.971571) and Office of
Naval Research is acknowledged. J.G.O. also acknowledges support from
DGES-Spain (projects PB96-0241 and PB98-0935). Spectrogram provided
by the Center for Spoken Language Understanding, Oregon Graduate
Institute of Science and Technology. Two-dimensional image provided
by A+ Free Clip Art.

}
\end{multicols}

\end{document}